\documentclass[twocolumn,aps,prb,showpacs,superscriptaddress,amsfonts,amssymb,amsmath]{revtex4}

\begin{document}
\draft
\title{Conservation of spin current}

\author{Jian Wang $^*$}
\address{Department of Physics and the center of theoretical and
computational physics, The University of Hong Kong,
Pokfulam Road, Hong Kong, China\\}
\author{Baigeng Wang}
\address{Department of Physics, Nanjing University, Nanjing, P.R. China\\}
\author{Wei Ren}
\address{Department of Physics and the center of theoretical and
computational physics, The University of Hong Kong,
Pokfulam Road, Hong Kong, China\\}
\author{Hong Guo}
\address{Department of Physics, McGill University, Montreal, Quebec, Canada
\\}

\begin{abstract}
The conventional definition of spin-current, namely spin density
multiplied by the group velocity, is not a conserved quantity due to
possible spin rotations caused by spin-orbit (SO) interaction.
However, in a model with spin-spin interactions, rotation of a spin
causes a dynamic response of surrounding spins that opposes the
rotation. Such a many-body effect restores the spin-current
conservation. Here we prove that the non-conservation problem of
spin-current can be resolved if a self-consistent spin-spin
interaction is included in the analysis.
We further derive a spin-conductance formula which partitions
spin-current into different leads of a multi-lead conductor.

\end{abstract}


\maketitle
Recently, considerable interest has been paid to the quantum physics of
spin-current\cite{sarma-review}. It is believed that a controlled
spin-current generation, detection, and usage can provide interesting
applications to spintronics. Spin-current generation has been classified as
``extrinsic'' or ``intrinsic''. An extrinsic spin-current is generated by
external physical factors and driving forces of the spintronics device, such
as optical spin injection achieved experimentally\cite{sipe} and the various
spin pumps studied theoretically\cite{sun-prl}. An intrinsic spin-current
is generated by physical factors existed inside the spintronic device,
notable is that generated by various spin-orbit
interactions\cite{das,zhang,sinova,kato,shen}.
In particular, it has been theoretically predicted that non-magnetic systems
with spin-orbit interaction and under an external electric field, can generate
a spin-current flowing perpendicular to the electric field\cite{zhang}. Such
an spin-current is termed ``dissipation-less'' because the electron motion
is perpendicular to the electric field. There are so far extensive
theoretical work on spin-current physics\cite{halperin,ref7}, and some experimental
works have also appeared\cite{kato,marcus1} which may provide support to
some of the theories.

Despite the increasing literature on spin-current physics, it is recognized that
the definition of spin-current itself is still somewhat controversial. If one
mimics the definition of charge-current, then a spin-current $I_s$ can be defined
as the time derivative of spin density $N_s$, $I_s\sim <dN_s/dt> $, here the bracket
is the quantum average. In its simplest form at steady state, such a definition gives
a spin-current $I_s=\frac{\hbar}{2}(I_\uparrow - I_\downarrow)$, where
$I_\uparrow, I_\downarrow$ are the charge current with spin-up and spin-down,
respectively. Clearly, this is a very intuitive definition of spin-current and
is adopted by most of the work in literature. Since both spin and velocity are
vectors, the spin-current is a tensor. In systems where there is a spin-orbit
interaction, the spin density is not conserved: spins can rotate from their
initial orientation due to the interaction. Therefore, spin-current $I_s$
becomes a non-conservative quantity. A quantity which is not conserved is
difficult to study experimentally and indeed, it is unclear what is even
measured if an experimental detection method can be found to measure
spin-current. Without knowing what is measured, the definition of spin-current
becomes non-unique and there have been several definitions in literature.

Although it is unclear {\it a priori} if a measurable spin-current
must be conserved, the property of conservation would be nice to
have, at least theoretically. Such an issue has been discussed in a
recent paper of Sun and Xie\cite{sun1} who defined an additional
rotational contribution to spin-current on top of the above
conventional definition, and the total spin-current is then
conserved. However, there is no microscopic theory concerning this
problem. In the present work, we further investigate this problem by 
using the conventional definition of spin-current $I_s\sim <dN_s/dt>$, 
but we include a spin-spin interaction into the Hamiltonian for computing 
the quantum average $<\cdots>$.  Our basic idea is the following. The 
existing problem of spin-current non-conservation was due to SO interaction
which rotates the spin away from its original direction. In a spin system
with spin-spin interactions, the rotation of one spin causes a
dynamic response of surrounding spins through the interaction, and
this response opposes the original spin rotation. This many-body
effect introduces a new term in the spin-current that balances the
spin-current non-conservation. In the end, one obtains a final
spin-current in the interacting model (which has a different value
than that of non-interacting model) that is conserved. In the
following, we demonstrate this idea by proving that $I_s$ obtained
in a model with dipole-dipole interaction is, indeed, conserved.
Importantly, the idea is based on a many-body phenomenon, as long as
there is spin-spin interaction, our conclusion should be general.
Furthermore, since there are always some dipole interactions between
spins, including their contribution is also rather natural. Finally,
the conserved spin-current gives a linear spin-conductance, we
derive an expression for this spin-conductance for multi-probe
device systems.

We consider a coherent multi-probe mesoscopic device which consists a device
scattering region that connects to a number of leads extending to far away. In a
lead labeled by $\alpha$, the spin-current
$I_{s\alpha} \sim <dN_s/dt>$.  Following Ref.\cite{jauho}, we calculate the
time derivative using Heisenberg equation of motion and in steady state, the spin
current $I_s$ can be written in terms of Green's functions:
\begin{eqnarray}
I_{s \alpha} &=&-\int \frac{dE}{4\pi } {\rm Tr}
\left[ \sigma_3 ({\bf G}^r {\bf \Sigma}^<_\alpha -  {\bf \Sigma}^<_\alpha
{\bf G}^a \right.\nonumber \\
&+&\left.{\bf G}^< {\bf \Sigma}^a_\alpha -{\bf \Sigma}^r_\alpha {\bf G}^<)
\right]
\label{scur1}
\end{eqnarray}
where ${\bf \Sigma}^{r,a}$ are the retarded and advanced self-energies due to
the presence of leads; ${\bf G}^{<,r,a}$ are the lesser, retarded and advanced
Green's functions of the device scattering region, respectively. Note that the
trace in the last equation can be reduced as:
\begin{eqnarray}
{\rm Tr} \left[ \sigma_3 ({\bf G}^r {\bf \Sigma}^< -
{\bf \Sigma}^< {\bf G}^a )\right] 
={\rm Tr} [{\bf G}^a F {\bf G}^r {\bf \Sigma}^<_\alpha]
\nonumber
\end{eqnarray}
where
\begin{eqnarray}
F \equiv [{\bf G}^{a}]^{-1} \sigma_3 - \sigma_3 [{\bf G}^{r}]^{-1}
= [\sigma_3,H_0] -{\bf \Sigma}^{a} \sigma_3 - \sigma_3 {\bf \Sigma}^{r}
\ .
\nonumber
\end{eqnarray}
Here $H_0$ is the Hamiltonian of the device scattering region without the
leads. Hence Eq.(\ref{scur1}) can be written as
\begin{eqnarray}
I_{s \alpha} &=&-\int \frac{dE}{4\pi } {\rm Tr}
\left[ {\bf G}^< ({\bf \Sigma}^{a}_\alpha \sigma_3 - \sigma_3
{\bf \Sigma}^r_\alpha) \right.\nonumber \\
&-&\left.{\bf G}^<_\alpha ({\bf \Sigma}^{a} \sigma_3 - \sigma_3
{\bf \Sigma}^{r}) + [\sigma_3,H_0] {\bf G}^<_\alpha \right]
\label{scur2}
\end{eqnarray}
where ${\bf G}^<_\alpha = i{\bf G}^r {\bf \Gamma}_\alpha f_\alpha {\bf
G}^a$ with $\Gamma_\beta$ the linewidth function.

The total spin-current is
\begin{eqnarray}
I_s\ =\ \sum_\alpha I_{s\alpha} = -\frac{\partial S}{\partial t}
\label{conti}
\end{eqnarray}
where
\begin{eqnarray}
\frac{\partial S}{\partial t} \equiv
i\int \frac{dE}{4\pi } {\rm Tr} [{\dot \sigma_3 } G^<]\ \ .
\label{accu}
\end{eqnarray}
with ${\dot \sigma_3} = -i [\sigma_3, H_o]$. Eq.(\ref{conti}) is just the
continuity equation for spin current in the absence of external magnetic
field and spin relaxation\cite{stiles}. The physics of this equation is
rather clear: in the presence of any spin-orbit interaction the electron spin
precesses due to an internal torque on the spin, hence the total spin current
flowing into the scattering region is non-zero and it equals to the rate of spin
precession $-\partial S/\partial t$. This is as if there were some ``spin accumulation"
in the scattering region. In other words, 
the total spin-current $I_s$ is itself not conserved due to spin precessing.

This situation is reminiscent to the continuity equation of {\it charge current}
$I_e$ in the presence of a time dependent field\cite{buttiker}, namely
\begin{eqnarray}
I_e\equiv \sum_\alpha I_{e,\alpha} = -\frac{\partial Q}{\partial t}
\label{conti-charge}
\end{eqnarray}
Here the total charge current $I_e$ is equal to the charge accumulation in the
scattering region.  The problem of charge current conservation under AC fields was
discussed by B\"uttiker\cite{buttiker} who pointed out that the total particle
current is not conserved under AC conditions, but the total particle current plus total
displacement current is a conserved quantity. The total displacement current is
precisely $\partial Q/\partial t$ in the above equation. Since displacement current
results from induction which is related to electron-electron interaction, B\"uttiker
formulated a current conserving theory\cite{buttiker} by including the electron-electron
interaction at the mean field level, which naturally deduces the displacement current in
each lead. The problem of displacement current partition in multi-probe conductor
under nonequilibrium conditions has been reported in Ref.\onlinecite{wbg}
within the nonequilibrium Green's function formalism.

The similarity between Eqs.(\ref{conti}) and (\ref{conti-charge})
suggests that the problem of spin-current conservation can also be
looked at from an interacting spin point of view.
As discussed above, a spin-spin interaction causes a dynamic
response inthe system which counter-react on any rotation of a spin.
In the following, we demonstrate this many-body physics by
investigating the consequence of a self-consistent spin-spin
interaction in a device Hamiltonian which contains the Rashba
spin-orbit (SO) interaction.  Indeed, we prove that the
self-consistent spin-spin interaction produces a term that exactly
cancels the right hand side of Eq.(\ref{conti}) so that $\sum_\alpha
I_{s\alpha} = 0$.

Our model Hamiltonian is ($\hbar=1$):
\begin{eqnarray}
H = H_o + V^{ss}
\nonumber 
\end{eqnarray}
where
\begin{eqnarray}
H_o = \sum_n (\epsilon_n +qU_n) d_n^\dagger d_n
+\sum_{nm} ( V^{so}_{nm} d_n^\dagger d_m
+h.c.)
\label{ham}
\end{eqnarray}
\begin{eqnarray}
H^{ss}= \frac{1}{2}\sum_{nmij} (V^{ss}_{nmij} d^\dagger_n d_m
d^\dagger_i d_j+h.c.)
\nonumber
\end{eqnarray}
where $H^{ss}$ is the spin-spin interaction and $n$,$m$,$i$, and $j$ include
spin indices. Here $V^{so}$ and $V^{ss}$, respectively, are matrix elements
of the Rashba SO interaction and the
spin-spin interaction. In real space, they are given by
\begin{eqnarray}
V^{so}(x) = \alpha_R {\bf \sigma} \cdot ({\hat z} \times {\bf k})
\end{eqnarray}
and\cite{jackson}
\begin{eqnarray}
V^{ss}(x,x') =  (g\mu_B)^2 {\bf \sigma} \cdot {\bf \nabla}
~ {\bf \sigma} \cdot {\bf \nabla'} \frac{1}{|x-x'|} \ .
\nonumber
\end{eqnarray}

To deal with transport problems, we make a mean field analysis on the
spin-spin interaction so that $H^{ss}$ becomes
\begin{eqnarray}
H^{ss} = \sum_{nm} V_{nm} d^\dagger_n d_m
\nonumber
\end{eqnarray}
with
\begin{eqnarray}
V_{nm} = \frac{1}{2}\sum_{ij} (V^{ss}_{nmij} <d^\dagger_i d_j> + h.c.)
\nonumber
\end{eqnarray}
In real space, $V_{nm}$ becomes,
\begin{eqnarray}
V(x) = g\mu_B {\bf \sigma} \cdot \nabla \phi_M =
- g\mu_B {\bf \sigma} \cdot {\bf H}
\label{vss}
\end{eqnarray}
with
\begin{eqnarray}
\phi_M = -\int_V \frac{\nabla' \cdot {\bf M}(x') }{|x-x'|} dx'
\label{phi}
\end{eqnarray}

The local magnetic moment ${\bf M}$ is determined by spin density which,
in the language of nonequilibrium Green's functions (NEGF), is given by
\begin{eqnarray}
{\bf M}(x) = -i g\mu_B \int \frac{dE}{2\pi} {\rm Tr}_s [ {\bf \sigma}
{\bf G}^<_{xx}]
\nonumber
\end{eqnarray}
where ${\rm Tr}_s$ is the trace over spin space and ${\bf G}^<_{xx}$ is the
diagonal element of ${\bf G}^<$. Note that ${\bf H} = -\nabla \phi_M$ is the
self-consistent effective magnetic field due to the spin-spin interaction.
The Green's function also depends on this effective field through $V(x)$,
\begin{eqnarray}
{\bf G}^r = \frac{1}{E-H_0- g\mu_B {\bf \sigma} \cdot \nabla \phi_M
-{\bf \Sigma}^r}\ .
\label{gr}
\end{eqnarray}
From Eq.(\ref{phi}), $\phi_M$ satisfies the following Poisson-like equation
\begin{eqnarray}
\nabla^2 \phi_M = -4\pi \rho_M
\label{poisson}
\end{eqnarray}
where $\rho_M = -\nabla \cdot {\bf M}$ is the effective density of
magnetic charge\cite{jackson}. These equations form a
self-consistent problem.

To solve Eq.(\ref{poisson}), we consider a very large volume of space
surrounding the device scattering region such that the total magnetic
charge inside that volume to be zero. Mathematically, this
consideration means:
\begin{eqnarray}
\int dx \nabla \cdot {\bf M}(x) = 0 ~~~~ {\rm or} ~~~~
\int \frac{dE}{2\pi} {\rm Tr} (\nabla \cdot {\bf \sigma} G^<) = 0
\label{condition}
\end{eqnarray}
where $ {\rm Tr} = {\rm Tr}_s {\rm Tr}_o$ includes the trace over both
spin space and orbit space. Eq.(\ref{condition}) can be achieved if
the volume in which we solve Eq.(\ref{poisson}) is so large that the effective
magnetic field ${\bf H}$ on the surface of that volume is zero.

Using the model Hamiltonian Eq.(\ref{ham}) and applying the Heisenberg
equation of motion, we can evaluate
\begin{eqnarray}
{\dot \sigma_3}\ =\ -i[\sigma_3,H_0] = 2i\alpha_R \nabla \cdot {\bf \sigma}\ .
\nonumber
\end{eqnarray}
This result together with Eq.(\ref{condition}) proves
\[
\int \frac{dE}{2\pi} {\rm Tr} ({\dot \sigma_3} G^<) = 0\ .
\]
Therefore the ``spin accumulation" of Eq.(\ref{accu}) is actually zero.
In other words, if the spin-spin interaction is included in the Green's function,
there will be no spin accumulation and the spin current is conserved.
We emphasize that in the self-consistent formalism,
Eqs.(\ref{scur1},\ref{gr},\ref{poisson}) form the basic set of equations
for the spin current conserving theory.

Now we derive a ``spin conductance'' ${\cal G}$ that corresponds to
the spin-current. When the external bias is small, we expand the
spin-spin interaction $V(x)$ in terms of bias $v_\alpha$,
$V(x) = \sum_\alpha u_\alpha v_\alpha$
where we have introduced the notion of spin dependent characteristic
potential\cite{buttiker,ma} $u_\alpha$ which satisfy the gauge invariant
condition
$\sum_\alpha u_\alpha=1$, i.e., the spin current depends only on the
difference in external bias. Expanding Eq.(\ref{scur2}) in terms of small
bias $v_\beta$, we find at zero temperature,
\begin{eqnarray}
I_{s\alpha} = \sum_\beta {\cal G}_{\alpha \beta} ~ v_\beta
\label{Is1}
\end{eqnarray}
where
\begin{eqnarray}
{\cal G}_{\alpha \beta} = {\rm Tr} [g_{\alpha \beta} - (\sum_\gamma
g_{\alpha \gamma})_{xx} u_\beta(x)]
\label{conduc}
\end{eqnarray}
where
the matrix $g_{\alpha \beta}$ is given by
\begin{eqnarray}
g_{\alpha \beta} &=& \frac{i}{4} \left[({\bf \Sigma}^{a}_\alpha \sigma_3 -
\sigma_3 {\bf \Sigma}^r_\alpha) -({\bf \Sigma}^{a} \sigma_3 -
\sigma_3 {\bf \Sigma}^r) \delta_{\alpha \beta} \right.
\nonumber \\
&&\left.  + [\sigma_3,H_0] \delta_{\alpha \beta}
\right] {\bf G}^r \Gamma_\beta {\bf G}^a
+ h.c.
\label{gab}
\end{eqnarray}
Note that the first term of Eq.(\ref{conduc}) comes from expansion of
Fermi distribution function and the second term involving the
characteristic potential is due to the spin-spin interaction term in
the expansion. Linearizing Eq.(\ref{poisson}) we have
\begin{eqnarray}
-\nabla^2 \phi_\alpha(x) =\kappa
{\rm Tr}_s [(\sum_{\eta\gamma} g_{\eta\gamma})_{xx} u_\alpha(x)
-(\sum_{\gamma} g_{\gamma\alpha})_{xx}]
\label{poi1}
\end{eqnarray}
where $\kappa=(2\pi g\mu_B)/\alpha_R$ and $\phi_M = \sum_\alpha
\phi_\alpha v_\alpha$. Note that the expansion over external bias, the spin
conductance $g_{\alpha \beta}$ in both Eq.(\ref{gab}) and Eq.(\ref{poi1})
do not depend on the self-consistent interaction. From Eq.(\ref{vss})
we have $u_\alpha = g\mu_B \sigma \cdot \nabla \phi_\alpha$. From Poisson
like equation Eq.(\ref{poi1}) we obtain the spin dependent characteristic
potential and the spin conductance can be calculated from Eq.(\ref{conduc}).

We emphasize that this conductance guarantees that the linear spin-current
of Eq.(\ref{Is1}) is conserved.  Without the spin-spin interaction,
the conductance would be given by only the first term on the right hand
side of Eq.(\ref{conduc1}), and the resulting spin-current would not be
conserved.  In fact, the sum of the second term of Eq.(\ref{conduc1})
over space, {\it i.e.},
the quantity ${\rm Tr}[\sum_{\gamma} g_{\gamma\alpha}]$, is exactly
equal to the "spin accumulation" $-\partial S_\alpha/\partial t$ in
the scattering region from lead $\alpha$ due to Rashba interaction in
the small bias limit.  To see this, we find from Eq.(\ref{gab})
\begin{eqnarray}
\frac{\partial S_\alpha}{\partial t} = -{\rm Tr}[\sum_{\gamma}
g_{\gamma\alpha}]
= -\frac{i}{2}{\rm Tr} \left[ [\sigma_3,H_0] {\bf G}^r \Gamma_\alpha
{\bf G}^a \right]
\nonumber
\end{eqnarray}
so that $\sum_\alpha \partial S_\alpha /\partial t ~ v_\alpha =
\partial S/\partial
t$. This means that the spin-spin interaction puts this contribution into
the spin conductance itself automatically, so that the right hand side of
Eq.(\ref{conti}) vanishes.  Eq.(\ref{conduc1}) therefore partitions the
non-conserving part of the spin-current ({\it i.e.} the right hand side
of Eq.(\ref{conti}) when there is no spin-spin interaction) into each
leads, such that the spin-current becomes conserved.

Since the solution of the Poisson like equation requires numerical
calculation, analytically we can avoid this by using a quasi-neutrality
approximation, {\it i.e.}, assuming that the effective density of
magnetic charge $\rho_M(x)=0$ so that the local magnetic moment
${\bf M}$ is independent of position. Then, we find the spin dependent
characteristic potential by setting the right hand side of
Eq.(\ref{poi1}) to zero and obtain:
$u_\alpha(x) = (\sum_{\gamma} g_{\gamma\alpha})_{xx}/
(\sum_{\eta\gamma} g_{\eta\gamma})_{xx}$.
The conductance is then found to be:
\begin{eqnarray}
{\cal G}_{\alpha \beta} = {\rm Tr} [g_{\alpha \beta} - \frac{
(\sum_\gamma g_{\alpha \gamma}) (\sum_{\gamma} g_{\gamma\alpha})}
{(\sum_{\eta\gamma} g_{\eta\gamma})} ]\ \ .
\label{conduc1}
\end{eqnarray}
This expression is very similar to that of charge current partition in
ac situations\cite{buttiker,wbg}, and it partitions the total spin-current
into each lead $\alpha$ so that the total spin-current flowing into
the device is conserved.

The above microscopic theory result, Eq.(\ref{conduc1}), is valid for
Rashba SO interaction. For a general SO
interaction, a similar expression to Eq.(\ref{conduc1}) can be derived
using a phenomenological argument\cite{buttiker}. To do that, we require
two conditions: (i) the total spin current is conserved; (ii) the
value of spin current depends only on the difference of external bias.
The latter condition means that spin current remains unchanged if
external bias at each lead is shifted by the same amount. Now, the
unconserved spin current $I^c_{s\alpha}$ is given by Eq.(\ref{Is1}).
The ``spin accumulation" $I^d_s \equiv \partial S/\partial t$ is given
by $\sum_\alpha I^c_{s\alpha} = \sum_\beta (\sum_\alpha {\cal G}^c_{\alpha
\beta}) v_\beta = - I^d_s$, where we have used ${\cal G}^c_{\alpha \beta}
\equiv {\rm Tr} g_{\alpha \beta}$ for the non-conserved spin conductance.
Note that the total ``spin accumulation" is due to the contribution
$I^d_{s\alpha}$ from each lead $\alpha$, {\it i.e.}
$I_s^d = \sum_\alpha I^d_{s\alpha}$. Since only the total ``spin
accumulation" is known, we need to find $I^d_{s\alpha}$ by partition the
spin current. For this purpose, let $I_{s\alpha} \equiv I^c_{s\alpha} +
A_\alpha I^d_s$, or equivalently
\begin{eqnarray}
{\cal G}_{\alpha \beta} = {\cal G}^c_{\alpha \beta} -A_\alpha \sum_\gamma
{\cal G}^c_{\gamma \beta}
\label{parti}
\end{eqnarray}
where $A_\alpha$ is an unknown to be determined.
Condition (i) gives $\sum_\alpha {\cal G}_{\alpha \beta} = 0$, hence
we obtain $\sum_\alpha A_\alpha=1$. Condition (ii) gives gauge invariance
$\sum_\beta {\cal G}_{\alpha \beta} = 0$, hence we obtain
${\rm Tr} [\sum_\beta g_{\alpha \beta} -A_\alpha \sum_{\gamma\beta}
g_{\gamma \beta}]=0$
from which we find $A_\alpha = \sum_\gamma {\cal G}_{\alpha \gamma}/
\sum_{\gamma\eta} {\cal G}_{\gamma \eta}$. Therefore Eq.(\ref{parti})
gives:
\begin{eqnarray}
{\cal G}_{\alpha \beta} = {\cal G}^c_{\alpha \beta} - \frac{
(\sum_\gamma {\cal G}^c_{\alpha \gamma}) (\sum_{\gamma}
{\cal G}^c_{\gamma\alpha})} {(\sum_{\eta\gamma}
{\cal G}^c_{\eta\gamma})} \ \ .
\label{conduc2}
\end{eqnarray}
Eq.(\ref{conduc2}) has the same form as Eq.(\ref{conduc1}) which is
specific to Rashba SO.
We therefore propose that Eq.(\ref{conduc1}) can serve as a
phenomenological theory which conserves spin-current regardless of the
detailed spin-orbit interactions. Namely, if we use Eq.(\ref{conduc1}) to
compute spin-conductance, the resulting spin-current from Eq.(\ref{Is1})
will always be conserved regardless of which SO interaction is present.

In summary, we have proven that the conventional spin-current
$I_s\sim <n_s v_s>$ for Rashba interaction becomes a conserved
quantity if spin-spin dipole interaction is included. Such a dipole
interaction introduces a self-consistent field which correlates
spins spatially. For general SO interactions, a phenomenological
theory for spin current partition is proposed which conserves spin
current, and the resulting spin-conductance has the same form as
that derived from the microscopic theory of Rashba interaction.

{\bf Acknowledgments.}
We gratefully acknowledge support by a RGC grant from the SAR Government
of Hong Kong under grant number HKU 7044/04P. B.G. W is supported by the
grant from NSFC under grant number 90303011 and H.G is supported by NSERC of
Canada, FQRNT of Qu\'{e}bec and CIAR.

\bigskip
\noindent{$^{*)}$ Electronic address: jianwang@hkusub.hku.hk}

\end{document}